# Mapping High-Temperature Superconductors
## A Scientometric Approach

Andreas Barth* and Werner Marx**

\* FIZ Karlsruhe, D-76012 Karlsruhe (Germany)
\*\* Max Planck Institute for Solid State Research, D-70569 Stuttgart (Germany)

## Abstract

This study has been carried out to analyze the research field of high-temperature superconductivity and to demonstrate the potential of modern databases and search systems for generating meta-information. The alkaline earth (A2) rare earth (RE) cuprate high-temperature superconductors as a typical inorganic compound family and the corresponding literature were analyzed by scientometric methods. The time-dependent overall number of articles and patents and of the publications related to specific compound subsets and subject categories are given. The data reveal a significant decrease of basic research activity in this research field. The A2 RE cuprate species covered by the CAS compound file were analyzed with respect to the occurrence of specific elements in order to visualize known and unknown substances and to identify characteristic patterns. The quaternary and quinternary cuprates were selected and the number of compound species as function of specific combinations of A2 and RE elements is given. The Cu/O and RE/A2 ratios of the quaternary cuprate species as function of A2 and RE atoms are shown. In addition, the research landscape of the $MgB_2$ related publications was established using STN AnaVist, an analysis tool recently developed by STN International.

## 1 Introduction

A new research discipline around high-temperature superconductors has been established by the unexpected discovery of the superconducting La-Ba-Cu-O system in the year 1986 by Bednorz and Müller [1]. Since that time, the number of alkaline earth (A2) rare earth (RE) copper oxides (A2 RE cuprates) as well as the amount of the related publications has strongly increased. Currently, there are about 30,000 species from this compound family being registered by Chemical Abstracts Service (CAS), and around 65,000 articles and patents dealing with these compounds have been published until present (see Table 1). Until now, the high transition temperatures of the A2 RE cuprates cannot be explained by means of the classical theory describing the former low-temperature superconductors. Even after two decades of comprehensive research, a convincing and satisfactory theoretical explanation is still in expectance [2].

The research activities dealing with superconductivity are comparable to the former research on semiconductors: A broad class of material has been investigated by the complete tool-set of experimental and theoretical solid-state physics. The large number of new superconducting compounds and related articles has brought about that scientists working within this research field have increasingly problems to overview their discipline and to stay up-to-date with the new literature. Even more problematic is the fact that it becomes more and more difficult to integrate the new work into the existing body of information and knowledge. On the other hand, modern information systems offer databases and analysis tools providing a better overview on the entire research field and thus facilitating to cope with the abundance of literature. However, due to lack of access and experience concerning suitable analysis tools and databases many experts do not take advantage of them.

The authors of this study point out that they are information specialists. They intend to demonstrate exciting possibilities of databases in chemistry and physics and of information mining based thereon in order to provide a different view from experimental or theoretical research on superconductivity. The literature and the compounds of the appropriate research field are based on the available databases and search systems (especially on the CAS files under STN International) and the data are analyzed by applying scientometric methods. It should be emphasized that the literature analysis of this study was focused on basic research in chemistry and physics and not on engineering and technology. The more applied the work is, the more incomplete is the coverage in literature databases. Applied work for example is sometimes not accessible because of the legitimate proprietary emphasis of the organizations involved.

## 2 Methodology

The databases offered by Chemical Abstracts Service (CAS), a division of the American Chemical Society (ACS), in combination with a highly sophisticated search system enables information specialists to select different kinds of meta-information which could be relevant for the research process. The CAS files and their search system are the most powerful and sophisticated source of substance related literature (either articles or patents) in the fields of chemistry, materials science and physics as well. The CAS literature file (CAPLUS) stretches back to 1900 and covers both articles and patents [3]. The CAS compound file (REGISTRY) contains all chemical species being mentioned within chemistry and related research fields [4]. These items (publications and compounds) are called documents or records. The search options and additional functions for carrying out statistical investigations which are available via the online service STN International [5] have made it possible to perform extensive scientometric studies. However, the competent use of such databases and search systems requires some experience and awareness of possibilities and pitfalls.

The CAS files make it possible to determine the time curves of publication subsets being related to specific compounds or compound families, i.e. compound systems with a specific combination of elements or periodic groups. The compound families are searched in the CAS Registry file and the resulting answer sets are then transferred to the CAS literature file. A subsequent search in the CAS literature file yields all publications of the original compound families. The publications may be restricted further by combining the original answer set of Registry numbers with additional search terms from the literature file. The methodological aspects of the compound analysis are discussed in detail below (see chapter 4.1). The number of publications can easily be plotted against the publication year. In Figures 1-4 the time curves of publications for the entire research field, for certain research topics, and for the papers related to specific superconducting compound families are shown.

## 3 Publication Analysis

### *3.1 Compound-specific Literature Analysis*

In order to provide a broad overview the development of the overall number of publications of the superconductivity basic research discipline has been determined. Figure 1 shows the time curves established by using the Physics Abstracts file (INSPEC) [6] and the CAS files (CAPLUS and REGISTRY) [3-4]. The INSPEC curves show (1) the number of publications with superconductivity as a title word or as a controlled term (CT = high-temperature superconductors, type I and type II superconductors, "supercond?" as truncated index term) and (2) the number of articles dealing with superconductivity as established under (1) combined with A2 cuprates (A2 elements in combination with copper and oxygen appearing in the element term (ET) search field. The CAS curve shows for comparison the number of publications with any A2 cuprates (not only the rare earth derivatives as in Figure 2a below) being registered by the CAS Registry file and dealing with superconductivity, i.e. linked with "supercond?" (truncated search term) as index terms (keywords). In contrast to INSPEC all material based patents are covered by the CAS literature file.

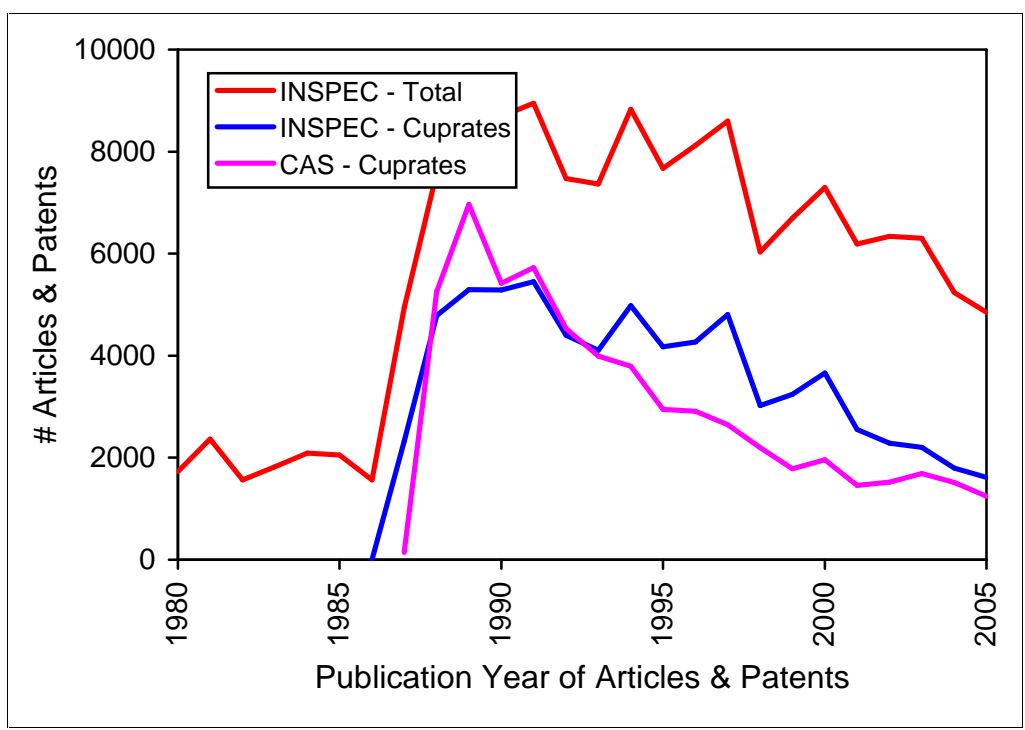

**Figure 1:** Time dependent number of publications related to superconductivity and covered by the literature file of Chemical Abstracts Service (CAS) and the Physics Abstracts file (INSPEC). The CAS data are cuprate based only. Source: CAPLUS and REGISTRY [3-4] and INSPEC on STN [6].

The striking increase of output is evidently originated by the discovery of the new high-temperature superconductors by Bednorz and Müller in the year 1986 [1]. Since around 1990 the overall number of publications (articles and material based patents) related to the entire field as covered by the databases consulted is decreasing significantly and reaching at present about half of the maximum. Please note that the overall scientific literature strongly increased within the last century and roughly doubled since the time when the new high-temperature superconductors were discovered [7]. In view of this increase the decrease of the superconductivity related publications is even more pronounced. Otherwise, the keyword based data related to the entire research field can not be established with the same precision (e.g. due to different indexing policies) as the succeeding material based output data.

The publications related to A2 cuprates containing RE elements and those related to A2 cuprates containing Bi, Tl or Hg instead of RE elements show a similar time evolution: The maximum of the publication activity occurs about four years after the discovery followed by a sustained decline of output (see Figures 2a). The output based on the given databases has decreased by about 70-80 percent with respect to the maximum output around 1989. The decreasing activity around the A2 cuprates containing Bi, Tl and Hg is somewhat astonishing, because these compounds turned out to score the highest transition temperatures until present. Please note that the data given here are based on the sophisticated CAS compound indexing. The publications were not further restricted by any keywords (e.g. superconductivity). For details concerning the procedure of searching see chapter 4.1 below.

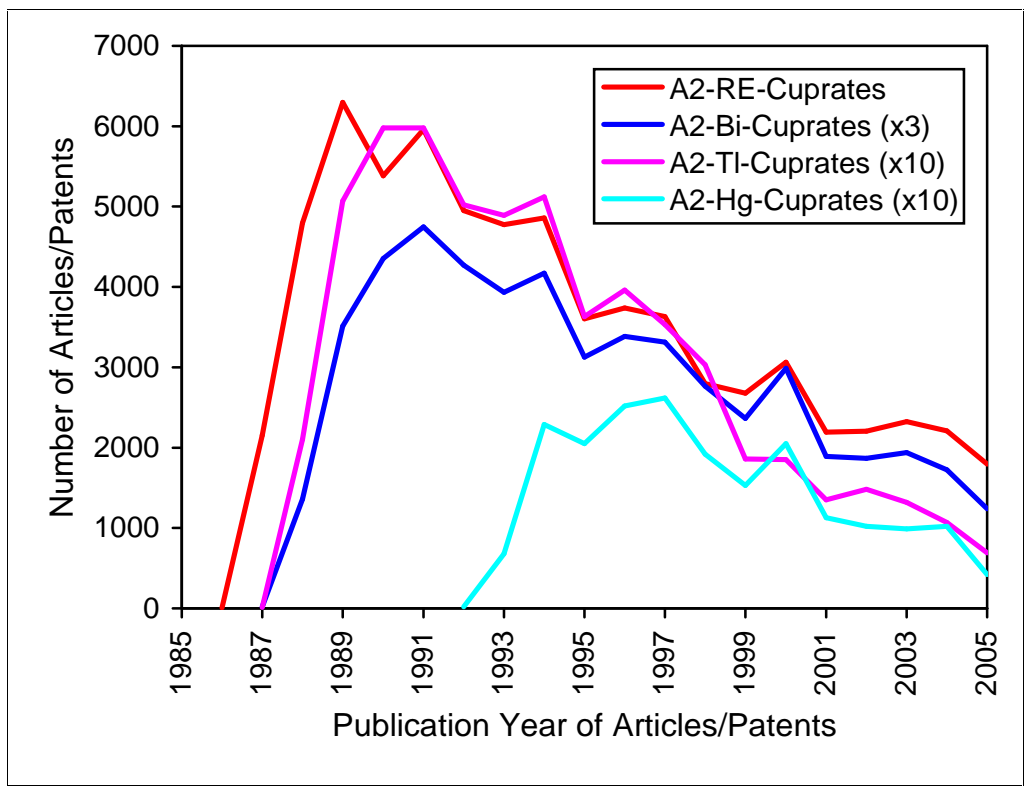

**Figure 2a:** Time dependent number of publications related to A2-RE cuprates and to A2 cuprates containing the elements Bi, Tl or Hg (not restricted to RE elements, number of different elements not limited). All material based patents are included. The papers were not further restricted by any keywords (e.g. superconductivity). Source: CAPLUS and REGISTRY on STN [3-4].

The time pattern of the magnesium boride ($MgB_2$) papers is given for comparison. $MgB_2$ is no high-temperature superconductor but a well known conventional material with a surprisingly high transition temperature, which was discovered not until the year 2001 and is thus of more recent interest. The data reveal that the number of publications dealing with MgB$_2$ decreased about one third within the last two years (see Figure 2b) which is quite similar compared to the cuprate related publications.

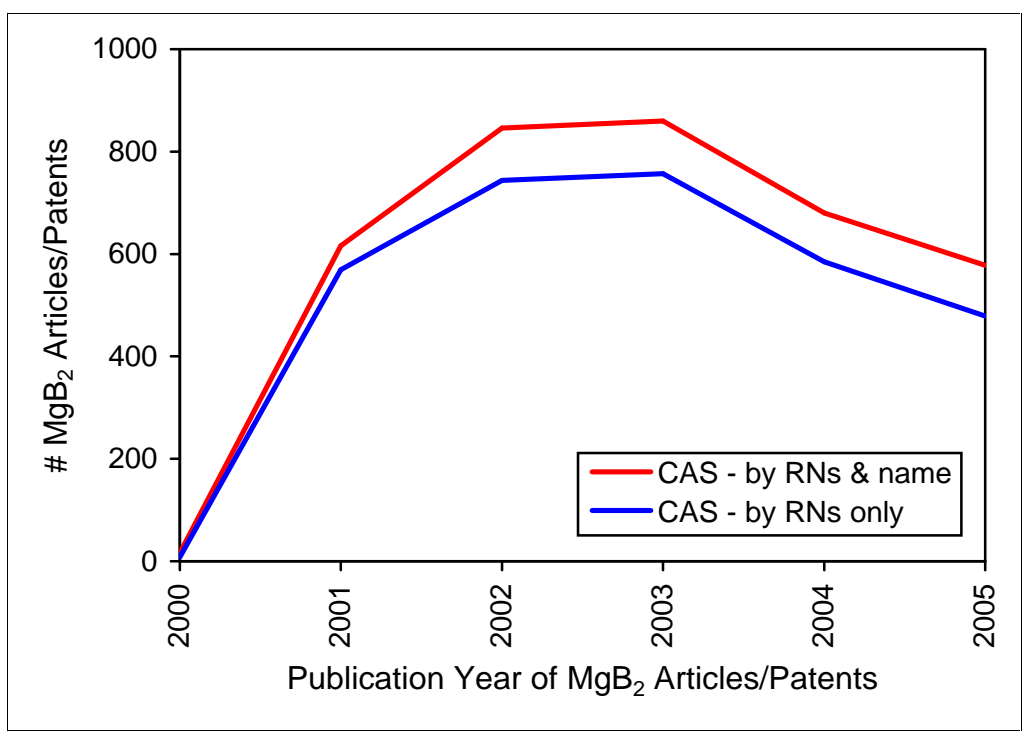

**Figure 2b:** Time dependent number of magnesium boride (MgB$_2$) publications. The literature covered by the CAS file has been searched by alternatively using only the MgB$_2$ chemical name and molecular formula or the CAS Registry number in addition. Source: CAPLUS and REGISTRY on STN [3-4].

Due to the absence of a satisfactory theoretical explanation of the phenomenon, a considerable amount of research can be expected also in the future. Groundbreaking discoveries like new superconductors with significantly higher transition temperatures or an unexpected superconducting material out of a new compound class would certainly boost the field again.

### 3.2 Keyword-specific Literature Analysis

The CAS indexing makes it possible to select and analyze publication sets dealing with superconductivity and related to specific categories. The complete set of A2 RE cuprates can be selected precisely in the compound file, then transferred to the literature file and finally combined with well defined keywords or roles. The index terms being added to the publications of the literature file contain the CAS Registry numbers of the compounds mentioned and in addition the related keywords (see chapter 4.1). Therefore, the publications discussing the preparation or the properties of A2 RE cuprates or dealing with theoretical investigations could be selected and analyzed. Figure 3 shows the time curves of the publications of these specific categories of investigation in superconductivity research.

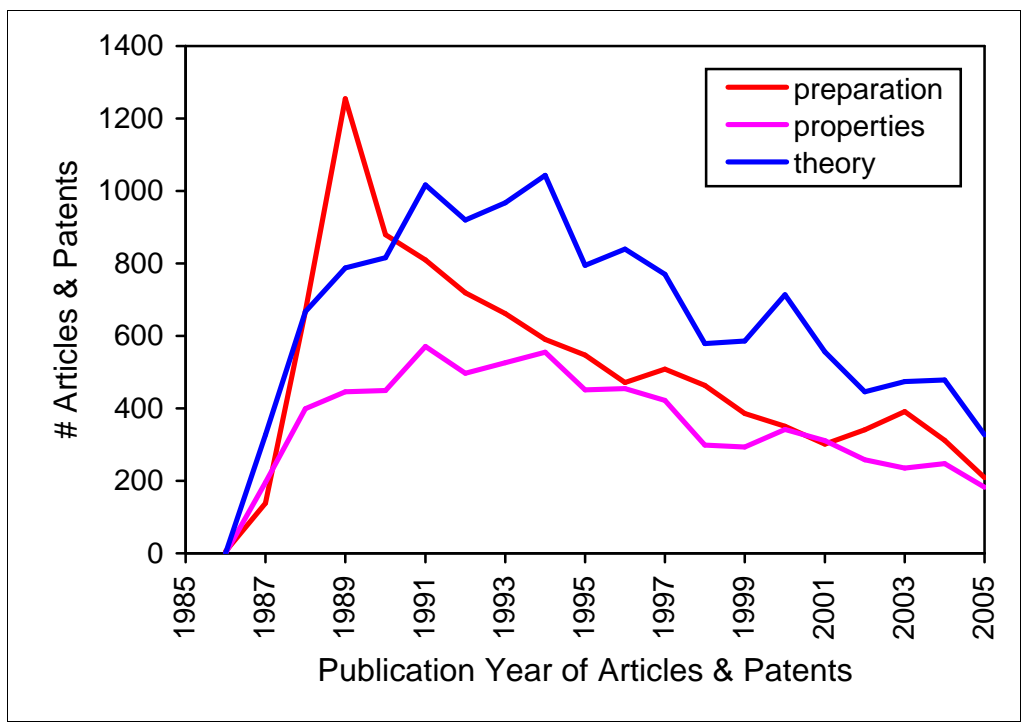

**Figure 3:** Time dependent number of articles related to the preparation, the properties, and to theoretical investigations of A2 RE cuprates. The compound family has been selected in the CAS Registry file and was then transferred to the CAS literature file. Source: CAPLUS and REGISTRY on STN [3-4].

The preparation related papers were selected by adding the qualifier "P" to the CAS Registry numbers. The property related papers were selected by the specific CAS role (PRP). The theory related papers were selected by combining the Registry numbers of the selected compounds with "calculat?", "simulat?", "model?" or "theor?" (truncated search terms) appearing in the title, the keywords or the abstract field.

The BCS theory, developed in 1957 by Bardeen, Cooper, and Schrieffer [9], was a breakthrough of theoretical solid state physics and it explains well the classical low-temperature superconductivity. The key idea is that electron phonon coupling results in an attractive interaction leading to bound electron pair states ("Cooper pairs"). Cooper pairs behave like bosons and move unresistingly through the crystal lattice. However, a new theoretical concept is required to explain the phenomenon of high-temperature superconductivity. It is speculated that magnetic interaction based on antiferromagnetism instead of phonon interaction originates Cooper pairs leading to high-temperature superconductivity [10]. This interesting hypothesis has to be verified by future research.

In order to visualize the ongoing discussion concerning the mechanism of high-temperature superconductivity, the time dependent number of articles dealing with "antiferromagnetism" in conjunction with superconductivity has been determined (see Figure 4). The Cooper pair related papers are shown for comparison. Please note that the antiferromagnetism related papers on superconductivity have been almost continuously published at a rate of around 300 articles per year.

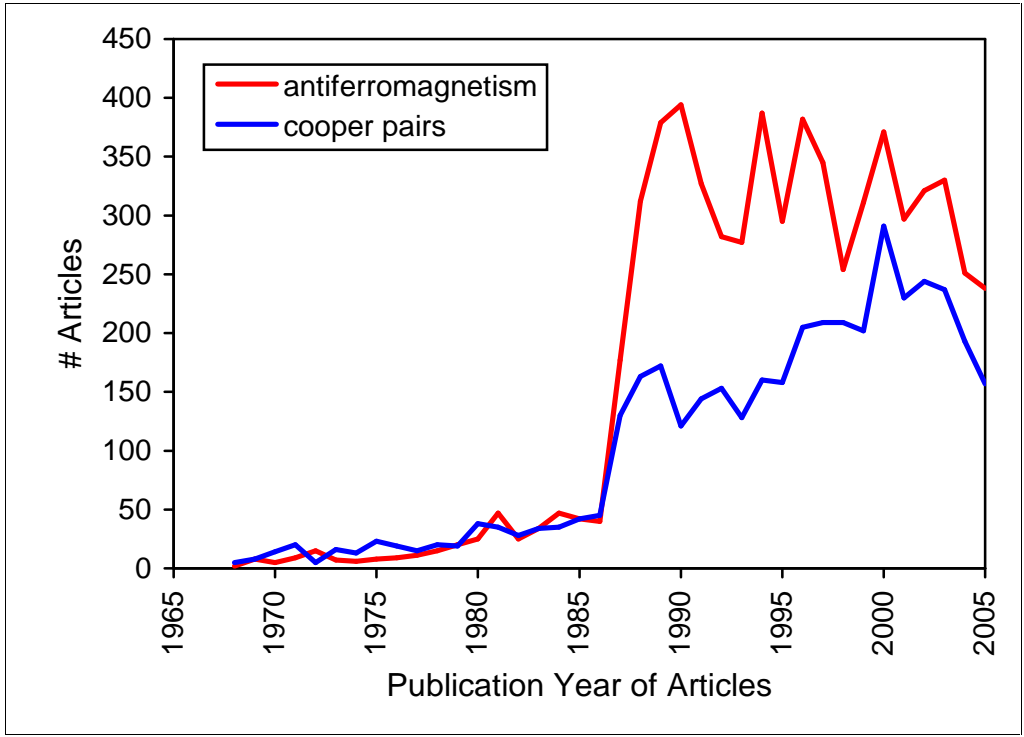

**Figure 4:** Time dependent number of articles mentioning "supercond?" in conjunction with "antiferromagnet?" and "cooper pairs" with the search terms appearing in titles, abstracts or keywords. Source: INSPEC on STN [6].

### 3.3 Research Landscapes

The online service STN International has recently launched a new interactive analysis tool called STN AnaVist [11] which focuses mainly on patent analysis [12]. This tool can also be applied to literature exploring the usefulness for basic research topics. One of the functions offered by STN AnaVist implies the creation of so-called research landscapes: "Significant keywords and concepts are derived from document titles and abstracts. These keywords are used to determine the similarity between documents. An algorithm uses document similarity scores to position each document relative to one another in a two-dimensional space, with each document positioned at one point. This process is repeated until all documents have been clustered and each assigned to a single x, y coordinate pair. A graphical map is generated. The z coordinates, determining the height of each 'peak', are calculated based on the density of the documents in an area."

As an example, the papers dealing with a single superconducting compound being currently of high interest were selected here: the magnesium boride ($MgB_2$) papers (see Figure 5). $MgB_2$ has been well known to Chemists since about half a century, but the discovery of superconductivity did not happen until 2001 quite by chance [13]. Unlike cuprate superconductors, $MgB_2$ seems to be a fairly conventional superconductor with an unexpectedly high transition temperature (almost 40 K) which is only exceeded by the more complicated cuprate materials. A search for publications of $MgB_2$ yields around 3000 papers until present. The $MgB_2$ landscape shows major studies around thin films (layers) and wires showing the very rapid progress in applying this material.

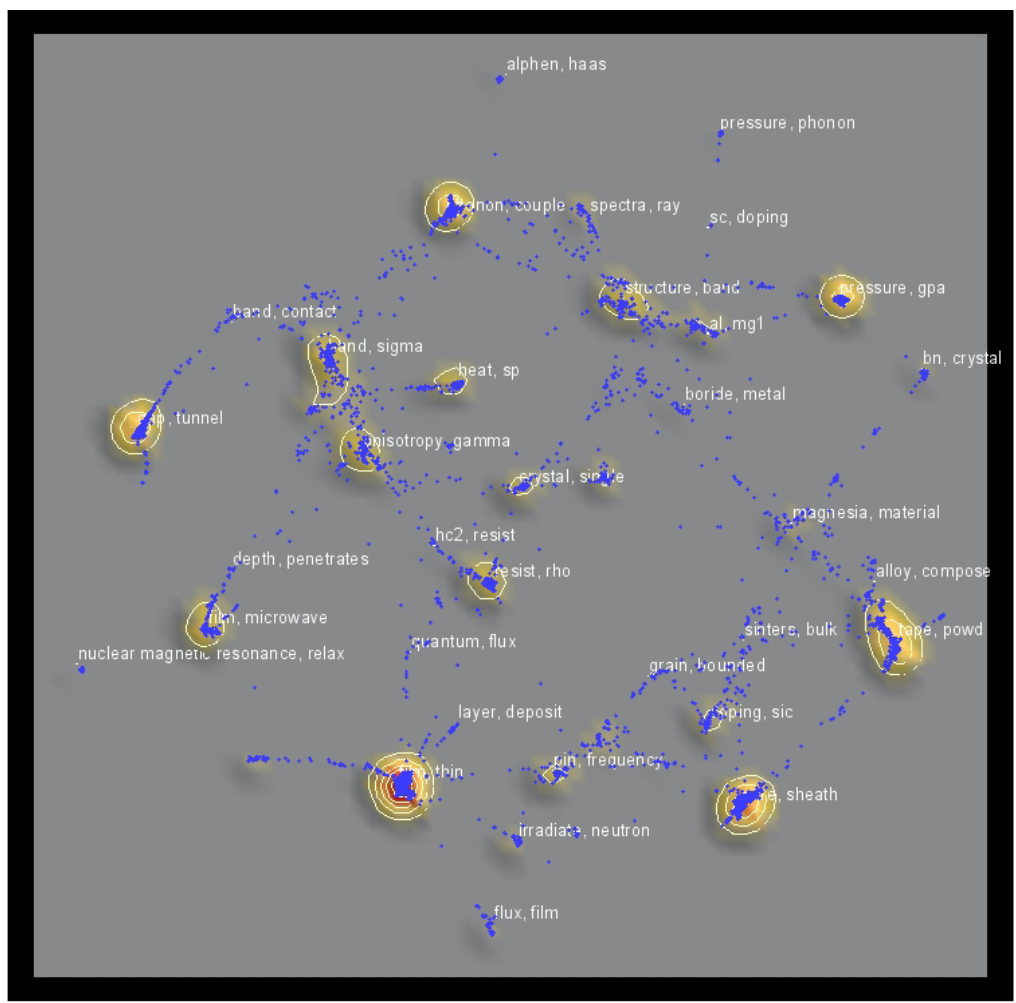

**Figure 5:** Research landscape of the MgB$_2$ publications established by the analysis tool STN AnaVist (STN International). Source: CAPLUS on STN [3].

## 4 Compound Analysis

### 4.1 Data Sources and Document Structure

The CAS Registry file implies two decisive advantages when searching chemical information: 1. Compounds are unambiguously coded by their CAS Registry numbers and 2. Compounds can be searched by using the sophisticated search possibilities of the Registry file (e.g. by searching their standard chemical names, molecular formulas, and structure formulas) instead of simply searching title or abstract words in the literature file - a highly incomplete and unspecific search method concerning compounds. As an example, the Registry file compound document (substance identification) of the earliest superconducting barium lanthanum copper oxide discovered by Bednorz and Müller [1] is shown (see Figure 6a).

```
L1    ANSWER 1 OF 1  REGISTRY  COPYRIGHT 2006 ACS on STN
RN    65107-47-3  REGISTRY
ED    Entered STN:  16 Nov 1984
CN    Barium copper lanthanum oxide (9CI)  (CA INDEX NAME)
OTHER NAMES:
CN    Copper barium lanthanum oxide
CN    Lanthanum barium copper oxide
DR    122097-43-2
MF    Ba . Cu . La . O
CI    TIS
LC    STN Files:  CA, CAPLUS, CIN, PIRA, TOXCENTER, USPAT2, USPATFULL

  Component      |      Ratio      |      Component
                 |                 |   Registry Number
 ==============+=================+==================
O                |       x         |     17778-80-2
Cu               |       x         |      7440-50-8
Ba               |       x         |      7440-39-3
La               |       x         |      7439-91-0

**PROPERTY DATA AVAILABLE IN THE 'PROP' FORMAT**

          365 REFERENCES IN FILE CA (1907 TO DATE)
            4 REFERENCES TO NON-SPECIFIC DERIVATIVES IN FILE CA
          365 REFERENCES IN FILE CAPLUS (1907 TO DATE)
```

**Figure 6a:** CAS Registry file compound document of the earliest superconducting barium lanthanum copper oxide. Source: REGISTRY on STN [4].

The Registry numbers can be transferred from the compound file to the literature file (and/or other CAS files) and vice versa. The available search functions of the STN search system make it possible to combine compounds (using the related Registry numbers) with index words in such a way that both the compounds and the keywords belong to the same index term (IT) and thus are interrelated to each other. As a consequence, publications related to specific compounds (or even large compound families like the A2 RE cuprates) in conjunction with specific research topics can be selected with high completeness and precision (i.e. high relevance). As an example for a CAS literature file document the famous paper by Bednorz and Müller [1], referring to the above lanthanum copper oxide, is given (see Figure 6b).

```
L1    ANSWER 1 OF 1  CAPLUS   COPYRIGHT 2006 ACS on STN
AN    1986:544390  CAPLUS <<LOGINID::20060517>>
DN    105:144390
TI    Possible high Tc superconductivity in the barium-lanthanum-
      copper-Oxygen system
AU    Bednorz, J. G.; Mueller, K. A.
CS    Zurich Res. Lab., IBM, Rueschlikon, Switz.
SO    Zeitschrift fuer Physik B: Condensed Matter (1986),64(2),189-93
      CODEN: ZPCMDN; ISSN: 0722-3277
DT    Journal
LA    English
IT    65107-47-3
      RL: PRP (Properties)
         (supercond. in)
```

**Figure 6b:** Bibliographic document of the famous paper by Bednorz and Müller [1], referring to lanthanum copper oxide (see Figure 6a). Source: CAPLUS on STN [3].

The CAS Registry file allows easy access to the complete set of compound species of the A2 RE cuprate compound family being registered until present. The selection is conducted here by combining the specific elements (Cu and O) and element groups (A2 and RE) possibly restricting to a specific number of different elements. The search query for the CAS Registry file is shown below (see Figure 6c).

```
=> dis hist

     FILE 'REGISTRY' ENTERED AT 17:14:11 ON 27 MAR 2006
CHARGED TO COST=IVS
L1     30495 S  CU/ELS(L)O/ELS(L)A2/PG(L)(B3 OR LNTH)/PG
L2      5967 S  CU/ELS(L)O/ELS(L)A2/PG(L)(B3 OR LNTH)/PG(L)4/ELC.SUB
```

**Figure 6c:** Search query of the search for A2 RE cuprates in the CAS Registry file. Source: REGISTRY on STN [4].

The A2 RE cuprate compound species are treated as individual compounds and thus are corresponding with a specific CAS Registry number (RN). The element combination is given in the molecular formula search field (MF), and the specific stoichiometry is shown in the alternative formula search field (AF). It must be emphasized here that differences in stoichiometry above 0.01 result in different compound species coded by specific CAS Registry numbers. As an example for a typical species, the complete $Ba_2YCu_3O_7$ compound document from the CAS Registry file is given (see Figure 6d). Please note the number of about 30,000 references in the literature file related to this compound (see further discussion in chapter 4.2).

```
L1    ANSWER 1 OF 1  REGISTRY   COPYRIGHT 2006 ACS on STN
RN    109064-29-1  REGISTRY
ED    Entered STN:  11 Jul 1987
CN    Barium copper yttrium oxide (Ba2Cu3YO7)(9CI)(CA INDEX NAME)
OTHER CA INDEX NAMES:
CN    Yttrate(4-), heptaoxotricuprate-, barium (1:2)
OTHER NAMES:
CN    Barium copper yttrium oxide (Ba4Cu6Y2O14)
CN    Barium yttrium copper oxide (Ba2YCu3O7)
CN    Barium yttrium cuprate (Ba2Cu3YO7)
CN    Copper barium yttrium oxide (Cu3Ba2YO7)
CN    SU 57
CN    YBCO
CN    Yttrium barium copper oxide (Ba2Cu3YO7)
CN    Yttrium barium copper oxide (YBa2Cu3O7)
CN    Yttrium barium cuprate (YBa2Cu3O7)
DR    866227-84-1
MF    Ba . Cu . O . Y
AF    Ba2 Cu3 O7 Y
CI    COM, TIS
SR    CA
LC    STN Files: CA, CAPLUS, CASREACT, CIN, PROMT, TOXCENTER, USPAT2,
      USPATFULL
   Component    |        Ratio        |      Component
                |                     |   Registry Number
==============+=====================+===================
O             |          7          |      17778-80-2
Y             |          1          |       7440-65-5
Cu            |          3          |       7440-50-8
Ba            |          2          |       7440-39-3

**PROPERTY DATA AVAILABLE IN 'PROP' FORMAT**

        30077 REFERENCES IN FILE CA (1907 TO DATE)
        21818 REFERENCES TO NON-SPECIFIC DERIVATIVES IN FILE CA
        30085 REFERENCES IN FILE CAPLUS (1907 TO DATE)
```

**Figure 6d:** CAS Registry file compound document of $Ba_2YCu_3O_7$. Source: REGISTRY on STN [4].

### 4.2 Coarse Analysis of Compound Families

The documents of the CAS compound file can be analyzed similar to the documents of the CAS literature file. Using the possibilities of element specific selection within the molecular formula search field (MF) and the alternative formula search field (AF) the number of cuprate compound species and the number of related articles covered by the CAS compound and literature files are given (see Table 1).

| system | # species | # articles | # element combin. |
|---|---|---|---|
| Cu - O | 211790 | 245862 | |
| Cu - O - A2 | 43068 | 93939 | |
| Cu - O - A2 - RE | 30495 | 64809 | |
| Cu - O - A2 - RE (4 Els) | 5967 | 57571 | 64 |
| Cu - O - A2 - RE (5 Els) | 15218 | 9842 | 1071 |
| Cu - O - A2 - RE (>5 Els) | 9306 | 3891 | 2560 |

**Table 1:** Number of cuprate compound species covered by the CAS Registry file. Source: REGISTRY on STN [4]. Date of search: March 27, 2006.

The number of different elements given in lines 1-3 of Table 1 is not limited (e.g. Cu - O implies all compounds containing copper and oxygen and any further element). The compound systems listed in lines 4-6 are limited to the number of different elements given in parenthesis - the number of element combinations is given for comparison. The overall number of articles (patents included) referring to the compounds covered by the CAS literature file (CAPLUS) is listed. Please note that the number of compound species is much higher than the pure number of element combinations: The quaternary system Cu-O-A2-RE, for example, comprises only 64 element combinations but includes 5967 compound species with different stoichiometry. One specific element combination like the quaternary system Y-Ba-Cu-O covers about 1500 compound species.

About 65,000 articles related to the A2 RE cuprate species were published almost completely later than 1985 - only 25 articles appeared before this year. Thus, the selected compound species (and the related publications) obviously deal almost exclusively with aspects of high-temperature superconductivity. The distribution of articles on the related superconducting compounds is highly skewed: a large fraction of articles referring to a small fraction of compounds. Out of almost 65,000 articles dealing with quaternary A2 RE cuprates, more than 30,000 articles deal with the specific compound $Ba_2YCu_3O_7$ and another 12,000 articles with the unspecific Ba-Y-Cu-O system. On the other hand, a large number of compound species of this quaternary element system is mentioned only a few times or even only once.

In addition to the variable composition of the compounds, different preparation methods and analysis accuracies cause a wide range of stoichiometry within one and the same compound family or compound system, repectively. Therefore, the number of different species of a specific element combination is more or less a measure of the amount of activity around this element composition - it is not a count of completely diverse compounds. Beside the cuprates there are other oxide based superconductors like oxobismuthates, the fullerides, and the carbides and borides. However, this study is mainly focused on the A2 RE cuprate compound family.

Compound families can be further analyzed in the CAS Registry file with respect to the occurrence of certain elements or element groups. The number of A2 RE cuprate species containing specific A2 and RE elements can be determined (see Figure 7). The cuprates were not restricted here to exactly four different elements - additional main group or transition metal group elements may be included. Again the number of compound species distinguished by the CAS Registry file is given.

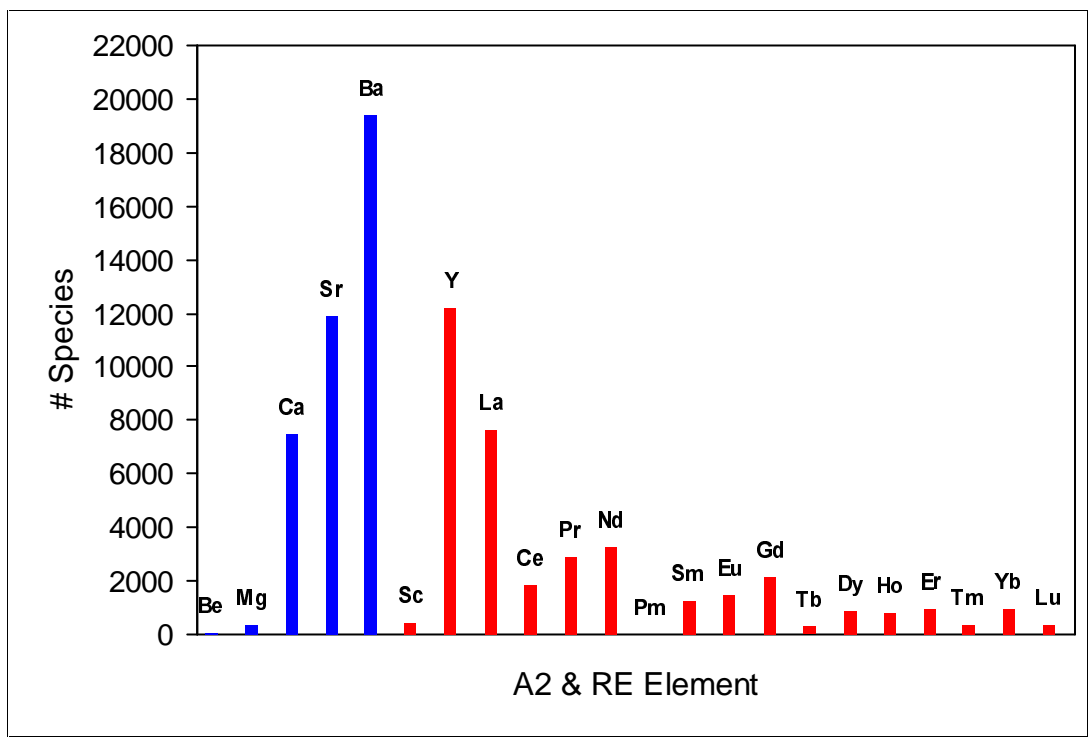

**Figure 7:** Number of A2 RE cuprate species covered by the CAS compound file and containing specific alkaline earth and rare earth (RE) elements. Source: REGISTRY on STN [4]. Date of search: March 7, 2006.

### *4.3 Compound Maps: 4 Elements*

More difficult to obtain are so-called compound maps which show the similar occurrence of two or more elements in a given compound family. In the case of quaternary cuprates the occurrence of A2 main group elements has been plotted against the rare earth (RE) elements (see Figures 8a-8b). Higher order compound maps with more than three elements are difficult to create and to visualize. However, it is possible to look at certain subsets and restrict the number of parameters for these compound maps. In this study we have created three-dimensional compound maps for quinternary barium and strontium cuprates (see Figures 9-11). The transition from 4 to 5 elements creates a broad family of compounds as shown in Figure 12 for the barium yttrium cuprates.

The compound families can be mapped to visualize the existing (and non-existing) species. All compound maps shown below were established using Microsoft Excel. As a first step the distribution of compound species within the quaternary A2 RE cuprate compound family has been analyzed by determining the number of species as function of specific combinations of A2 and RE elements. Out of 85 (5 x 17) possible element combinations of the quaternary system, 58 combinations were registered through the CAS compound file until present (see Figures 8a-8b). Figure 8a shows the view from the perspective of the A2 elements and Figure 8b shows the perspective of the RE elements. Date of search of the data presented in Figures 8-16: December 13, 2005.

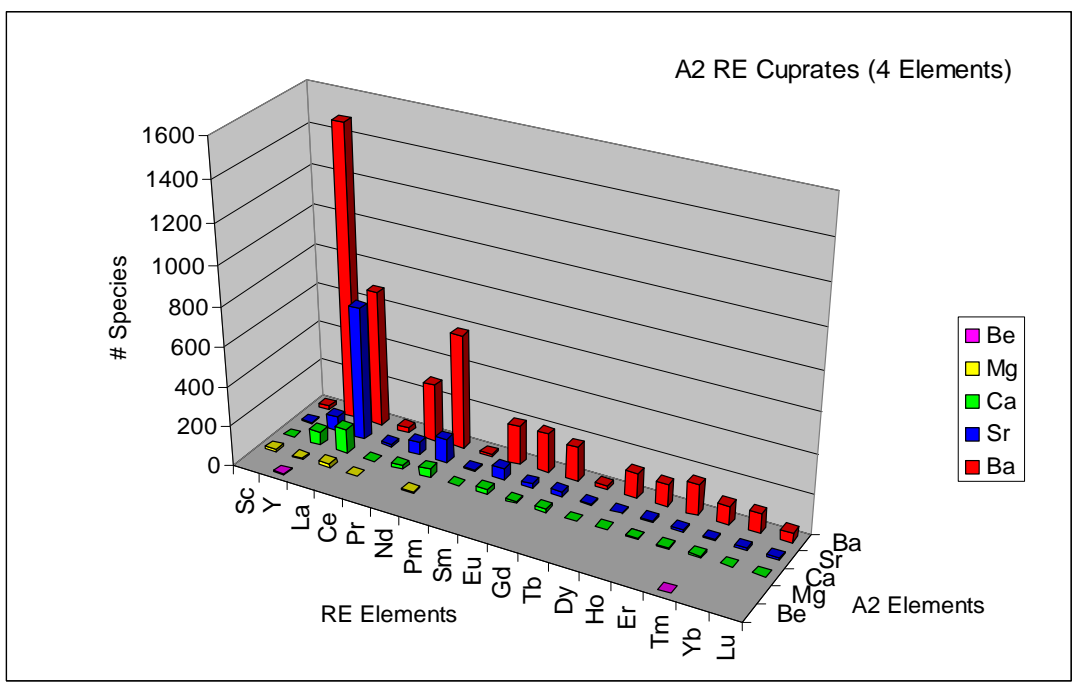

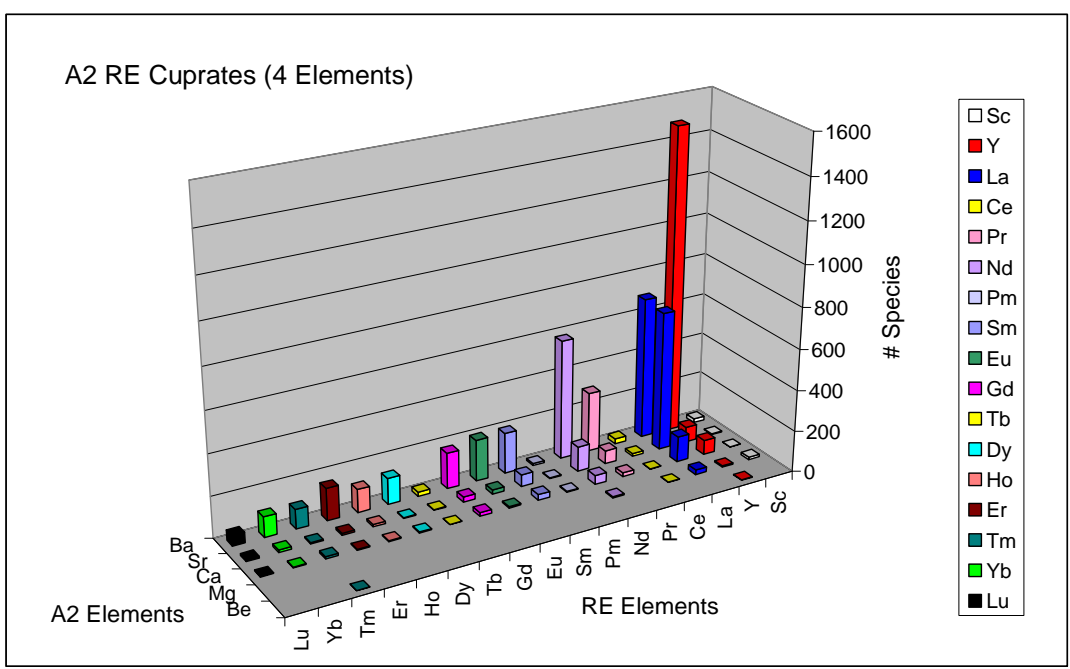

**Figures 8a-8b:** Number of compound species as function of specific combinations of A2 and RE elements of the quaternary A2 RE cuprate compound family. The two versions of color differentiation distinguish the possible A2 and RE compound subsets. Note: Two species with no AF (alternate formula) but with MF (Be-Tm-Cu-O, Mg-Ce-Cu-O) are included. Source: REGISTRY on STN [4].

### 4.4 Compound Maps: 5 Elements

Seen from the chemical point of view, superconductivity is a result of the interplay between chemical bonding or structure and the charge carriers. Superconductivity of A2 RE cuprates is believed to be originated by $CuO_2$ planes locating mobile charges. The metals and/or metal oxides located between the $CuO_2$ planes act as charge carrier reservoirs. A structure type showing superconductivity can be modified and optimized by substitution of the metal elements. Thereby, an enormous number of species has been generated, in particular the compounds with more than 4 elements.

Here, the barium and strontium RE cuprates containing exactly 5 different elements were analyzed concerning their element combinations (see Figures 9-10). Initially, compound species with two different RE elements were excluded. The 5th element, however, has been not further restricted. The CAS registration policy brings about that the addition of a fifth element with a stoichiometry above 0.01 generates a new quinternary compound species. However, such compounds may well be understood as quarternary compounds being doped by the fifth element.

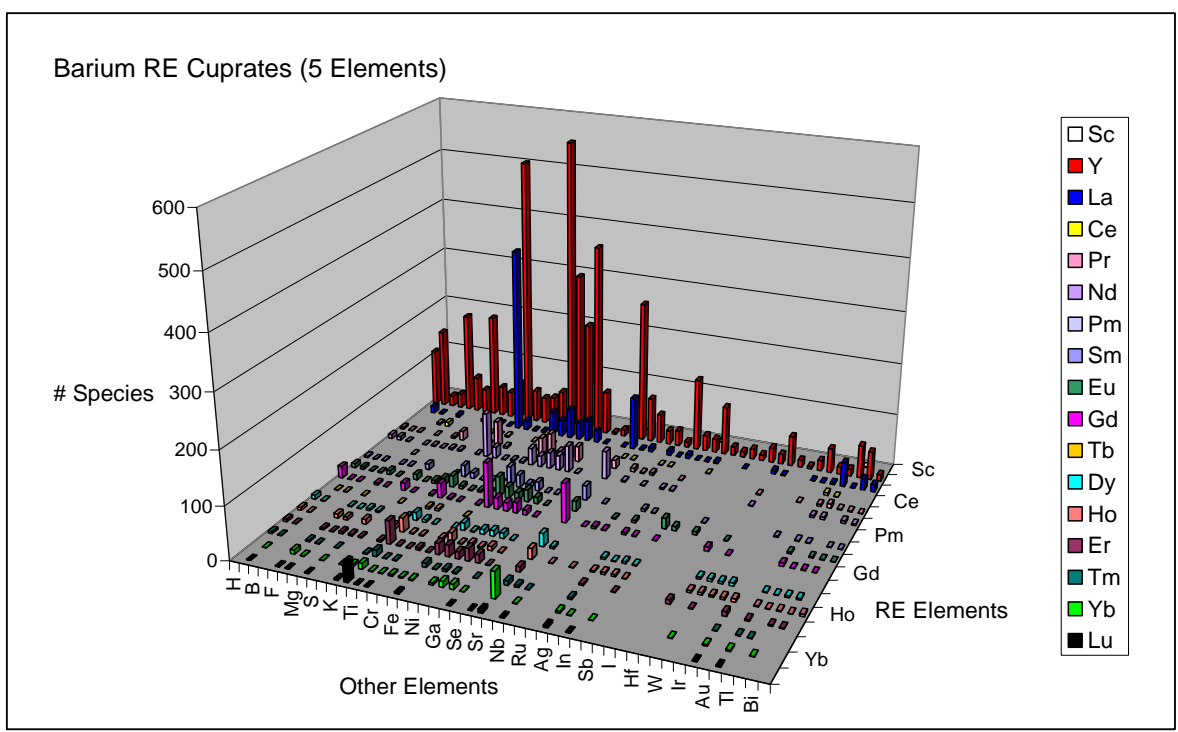

**Figure 9:** Number of compound species as function of specific combinations of RE elements with main group and transition metal elements of the quinternary RE barium cuprate compound family. Source: REGISTRY on STN [4].

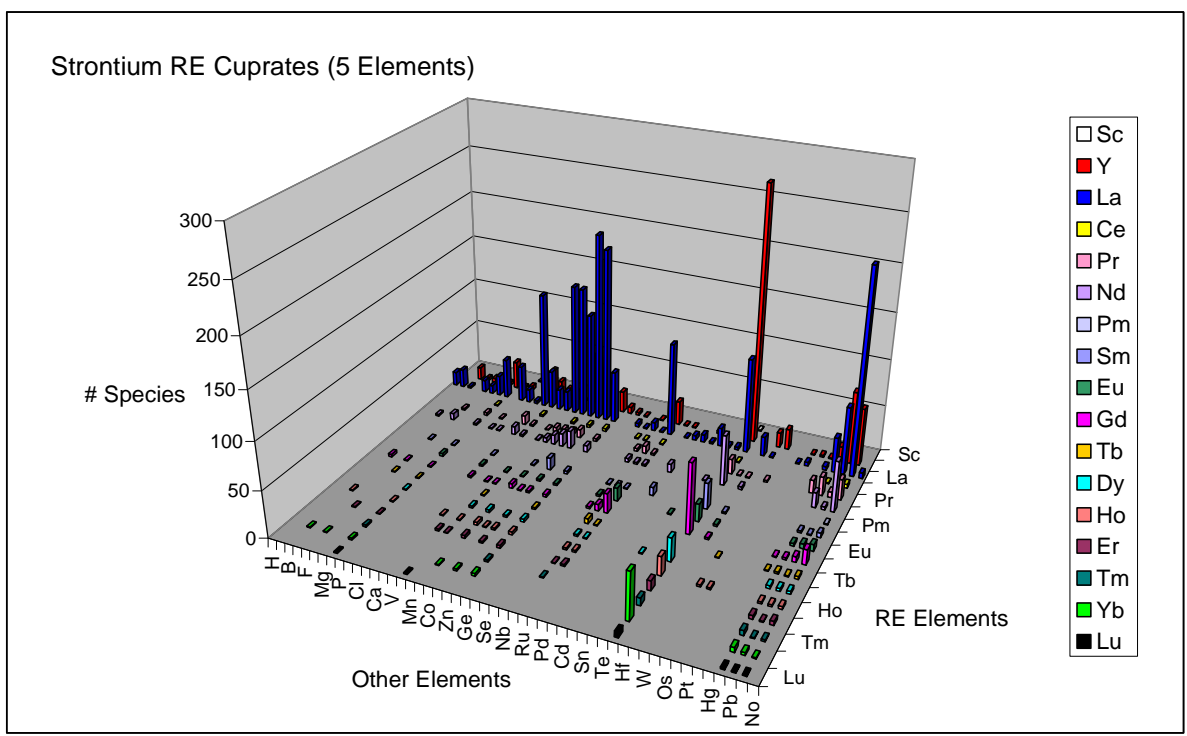

**Figure 10:** Number of compound species as function of specific combinations of RE elements with main group and transition metal elements of the quinternary RE strontium cuprate compound family. Source: REGISTRY on STN [4].

Apparently, the species containing yttrium are dominating the barium cuprates and lanthanum compounds are more numerous in the case of the strontium cuprates. This observation is also supported by the quaternary A2 RE cuprates (see Figures 8a-8b). In the case of barium RE cuprates there is only a single dominating peak for La (together with Ca). Similarly, in the case of strontium RE cuprates there is a single dominating peak for Y (together with Ba). Of course, the latter peak can also be seen in the diagram for barium RE cuprates with equal height. For completeness, barium cuprates with two different RE elements were analyzed separately (see Figure 11).

Please note that the RE element combinations have been doubled in order to generate a consistent symmetrical presentation. All peaks on the right hand side have a mirror peak on the left hand side. A look at the diagram from the right hand side shows all element combinations of a given RE element in a single color. It can be depicted from this diagram that the Y compounds in the back (in red color) are clearly dominating. The highest peaks in blue and pink color are actually mirror peaks of the most dominating red peaks for Y compounds.

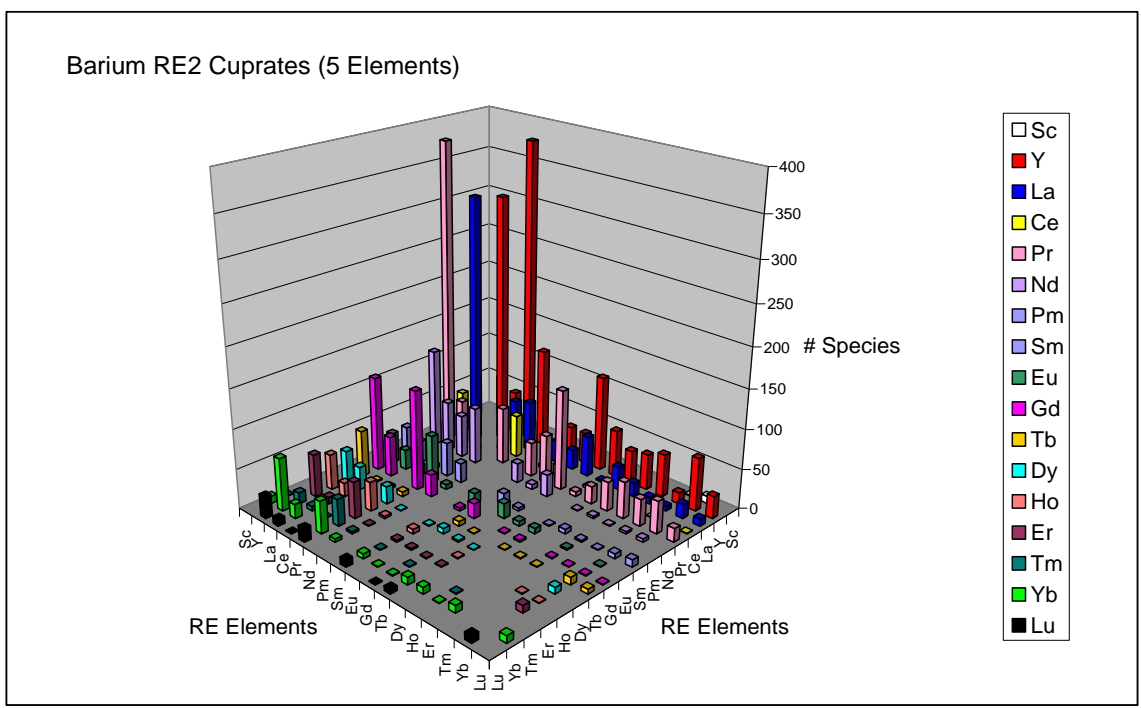

**Figure 11:** Number of compound species as function of specific combinations of RE elements among each other of the quinternary RE barium cuprate compound family containing two different RE elements. Source: REGISTRY on STN [4].

The transition from 4 to 5 elements can be visualized as a simple addition of a fifth element to a specific A2 RE element combination. This can be understood by looking at a simple example: In Figures 8a-8b the highest peaks show the number of quaternary Ba-Y cuprates. Adding another element X generates a series of compounds within the Ba-Y-X-Cu-O system. In principle X could be any other element but is actually limited to certain elements which have been selected for the preparation of new compounds. The distribution of these elements is contained in the graph of Figure 9 presented by the red peaks. To illustrate this transition from the Ba-Y-Cu-O system to a set of compounds with 5 elements the distribution of the 5th element is shown in Figure 12 in more detail.

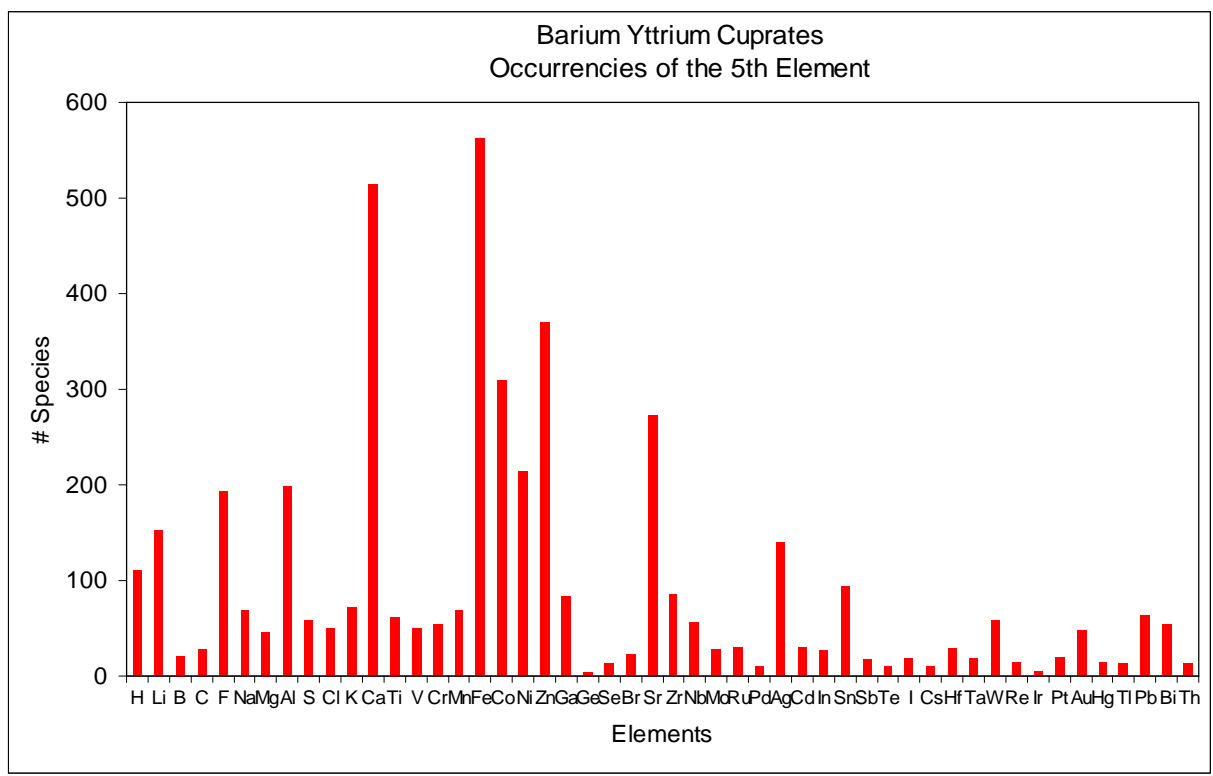

**Figure 12:** Distribution of the additional elements in quinternary Ba-Y cuprates (exactly 5 different elements). Source: REGISTRY on STN [4].

From a pure graphical point of view the outstanding single peak in Figures 8a-8b related to the barium yttrium cuprates has been expanded into the histogram in Figure 12. The number of compounds rises from 1,847 species of the Ba-Y-Cu-O system to 4,505 species of Ba-Y-X-Cu-O. Elements with highest frequencies are Fe (562), Ca (514), Zn (370), Co (309), and Sr (272). The distribution in Figure 12 indicates that barium is partly substituted by other elements.

### *4.5 Compound Maps: Cu/O Ratio*

Further analysis of element ratios of a given compound family is based on the stoichiometry given in the alternative formula search field (AF) of the Registry file. In the case of cuprates the Cu/O ratio is highly important. The oxygen content and thus the Cu/O ratio are strongly correlated with the transition temperature. In a first step the overall Cu/O ratio occurrence of the quaternary A2 RE cuprate species (exactly 4 different elements) has been analyzed. The number of species as function of the Cu/O ratio is shown (see Figure 13).

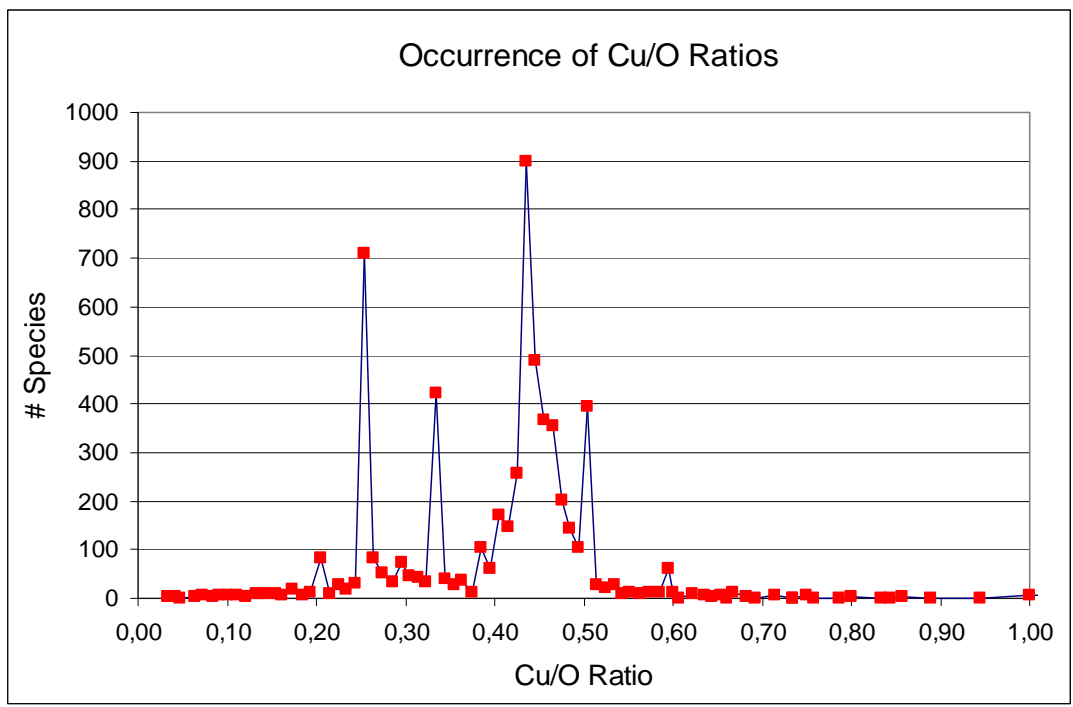

**Figure 13:** Cu/O ratio occurrence of the quaternary A2 RE cuprate species (exactly 4 different elements) covered by the CAS Registry file. The x-axis has been scaled to Cu/O ratios between 0.00 and 1.00 - two species with ratios of 1.33 and 5.26 were cut off. Source: REGISTRY on STN [4].

The Cu/O ratio ranges from 0.03 to 5.26 with an accumulation of higher occurrence in the range 0.40 - 0.50. Distinctive peaks of occurrence appear also at ratios of 0.25 (708 species), 0.33 (423 species), 0.43 (900 species), 0.44 (488 species), and 0.50 (396 species). Most of the species of the ratio 0.25 (600 out of 708) correspond to the $CuO_4$ compounds, most of the species of the ratio 0.33 (290 out of 423) correspond to the $CuO_3$ compounds, and most of the species of the ratio 0.43 (661 out of 900) correspond to the $Cu_3O_7$ compounds. Sometimes a stoichiometry range instead of a specific stoichiometry is given in the alternative formula search field (AF). For these compounds the mean values of the ranges were considered here.

Compound species with the highest transition temperatures like $HgBa_2Ca_2Cu_3O_8$ (133 Kelvin) and $Tl_2Ba_2Ca_2Cu_3O_{10}$ (125 Kelvin) correspond to Cu/O ratios of 0.37-0.38 and 0.30, respectively, thus being outside the ratios where the highest number of compound species occur. However, correlations between stoichiometry and transition temperatures are complicated due to a basic experience: The transition temperatures depend not only on the stoichiometry but largely on the crystal size, the purity of the compounds and their microstructure.

A more condensed view is obtained by mapping the occurrence of the specific elements out of the A2 and RE element groups against the Cu/O ratio (see Figure 14) or as function of the RE/A2 ratio (see Figure 15). Finally, the cation ratio (RE/A2) as a function of the Cu/O ratio has been determined (see Figure 16). The graphs of Figures 14-15 show an extract of the full diagram (the x-axis has been scaled to Cu/O ratios between 0.00 and 1.00) in order to demonstrate significant patterns. The element specific accumulation within the scaled Cu/O ratio range is visualized.

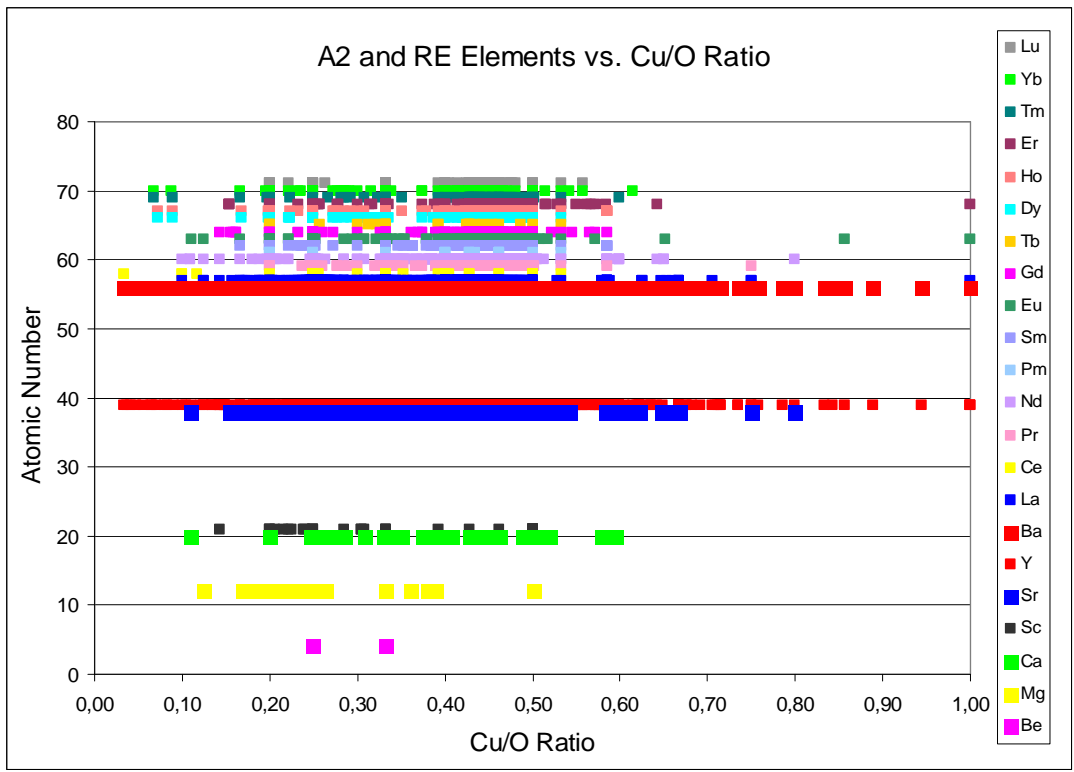

**Figure 14:** A2 and RE atoms as a function of the Cu/O ratio of the quaternary A2 RE cuprate species (exactly 4 different elements). The x-axis has been scaled to Cu/O ratios between 0.00 and 1.00 - two barium species with ratios of 1.33 and 5.26 were cut off. Source: REGISTRY on STN [4].

Each dot in this diagram denotes a compound species containing a specific A2 or RE element in combination with a certain Cu/O ratio. The element specific distribution (i.e. the accumulation and the vacancy pattern) is pronounced. Figure 14 shows a continuous range of Cu/O ratios between around 0.05 and 0.70 for the barium compound species corresponding to a similar range of the yttrium species. The other RE containing species are predominantly accumulated within a narrow range of Cu/O ratios between 0.40 and 0.60 with comparatively few species above this range. This is in accordance with the prevalence of the barium yttrium compound species as discussed above (see Figures 7-10).

It should be noted that the strong accumulation of species around certain ratios is not visible in this kind of graph. A three-dimensional graph with the number of species as a further parameter reveals that most of all species have Cu/O ratios within the range 0.20-0.50 with a distinctive maximum in the range 0.40-0.50. The number of species strongly decreases above 0.50. As an example, the variation of the number of barium containing species is given here: Cu/O ratio 0.20-0.30: 431 species, 0.30-0.40: 544 species, 0.40-0.50: 3043 species, 0.50-0.60: 127 species.

### 4.6 Compound Maps: RE/A2 Ratio

With respect to the cation stoichiometry, the RE/A2 ratio is the second important chemical degree of freedom. As with the Cu/O ratio, the ranges of stoichiometry were considered by processing the mean values. The ratio occurrence as function of the specific A2 and the RE elements has been determined. Again, the element specific accumulation and the range of the ratio are visualized (see Figure 15). The graphs show an extract of the full diagram (the x-axis has been scaled to RE/A2 ratios between 0 and 50) in order to demonstrate significant patterns. The element specific accumulation within the scaled RE/A2 ratio range is visualized.

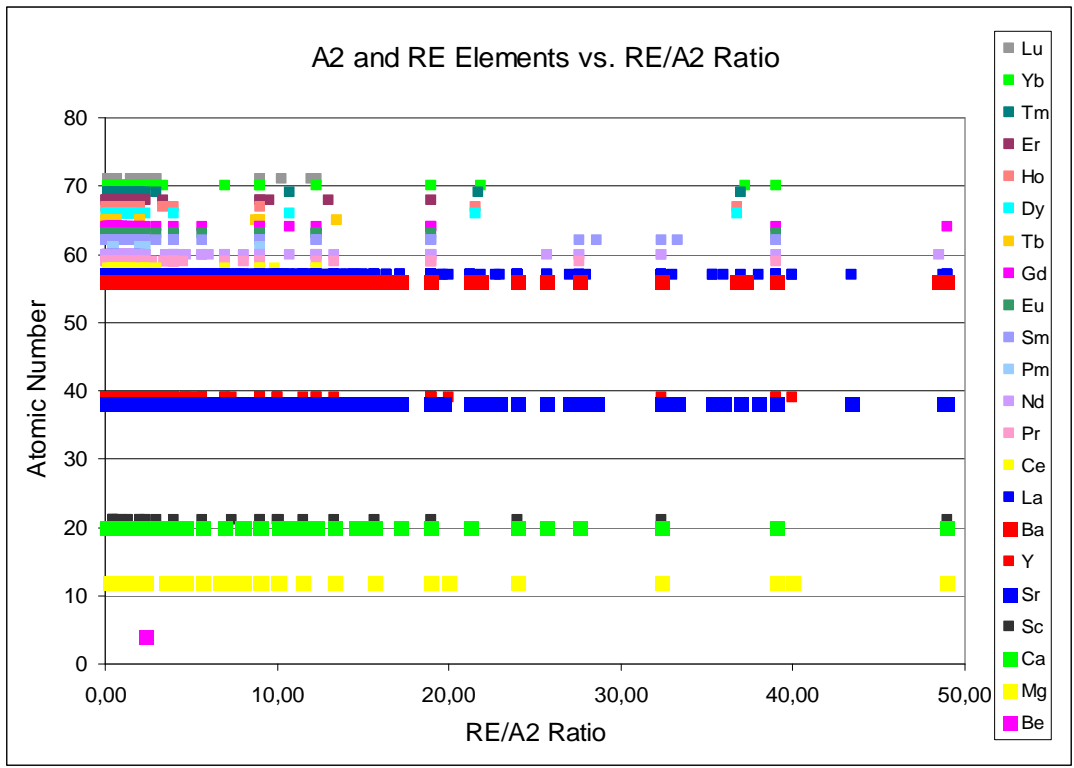

**Figure 15:** A2 and RE atoms as a function of the RE/A2 ratio of the quaternary A2 RE cuprate species (exactly 4 different elements). The x-axis has been scaled to RE/A2 ratios between 0 and 50 - 20 species with ratios up to 200 were cut off. Source: REGISTRY on STN [4].

Each dot in this diagram denotes a compound species containing a certain A2 or RE element in combination with a specific RE/A2 ratio. Again the accumulation of species and the vacancy pattern are pronounced. Figure 15 shows the wide range of ratios up to 50 - all species with ratios between 50 and 200 have been removed. It should be noted that the strong accumulation around a ratio of 1.00 and the preponderance of the ratio 0.50 as a result of the large number of $Ba_2YCu_3O_7$ derivatives are not visible in this graph. The higher ratios mainly result from species with an A2 stoichiometry of 0.01 (e.g. $Ba_{0.01}CuLa_{0.99}O_3$) and are continuously decreasing (with only 20 species out of almost 6000 showing ratios in the range between 50 and 200).

In order to show the direct relationship between the two degrees of freedom the cation ratio RE/A2 has been mapped against the Cu/O ratio (see Figure 16). The values higher than 1.00 for the Cu/O ratio and higher than 50.00 for the RE/A2 ratio have been cut off again, in order to demonstrate the combination pattern of these two parameters in more detail.

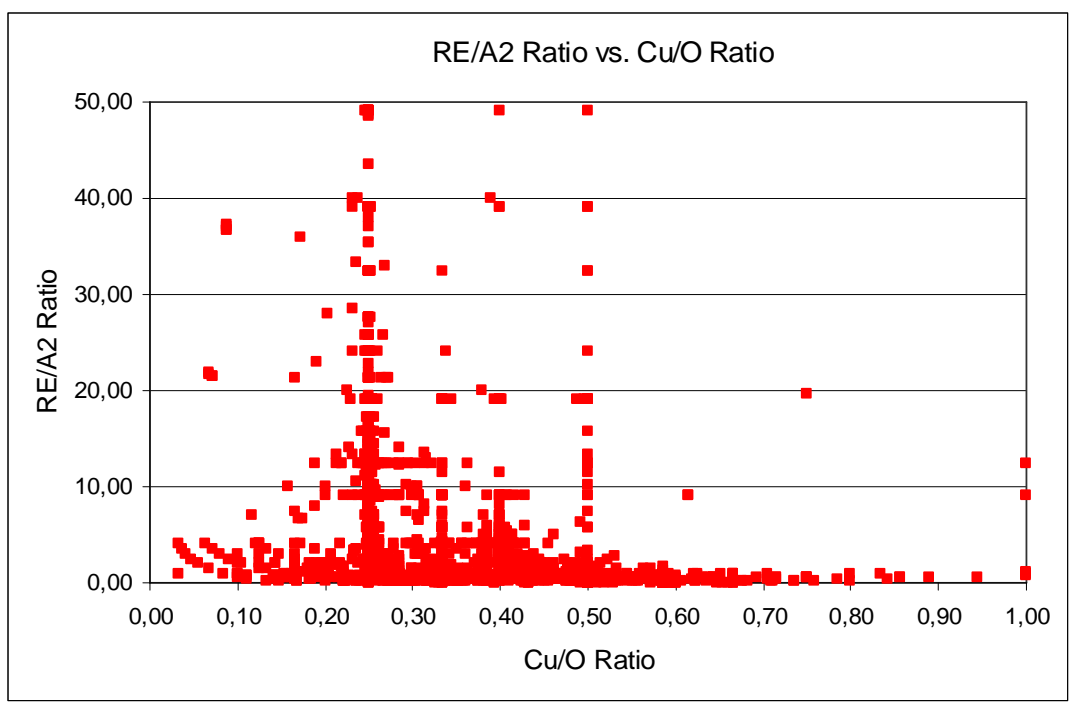

**Figure 16:** RE/A2 ratio as a function of the Cu/O ratio of the quaternary A2 RE cuprate species (exactly 4 different elements). Both axes were scaled according to Figures 14-15. Source: REGISTRY on STN [4].

Each dot in this diagram denotes a specific RE to A2 value at a certain Cu/O ratio. It should be noted that for a given value of the Cu/O ratio there may be more than one value of the RE/A2 cation ratio. The diagram shows a high concentration of low RE/A2 values almost along the complete Cu/O axis. The RE/A2 values show a large variation at certain values of Cu/O (e.g. at 0.20, 0.24, and 0.50).

**Summary**

A typical main stream research discipline has been reviewed by scientometric methods as a case study. The publications and compound species of the A2 RE cuprate high-temperature superconductors were analyzed and mapped. As a major consequence of the analysis the number of publications on superconductivity is strongly decaying. In addition to the literature analysis the A2 RE cuprate compound family has been investigated in detail with respect to several parameters. Compound maps have been generated to reveal known and unknown compounds together with the associated numbers of registered species. These maps enable researchers to detect white spots and to identify research potential for the preparation and analysis of unknown compounds. The analysis of the different parameters shows compound family patterns both with respect to the Cu/O ratio and the A2/RE cation ratio and demonstrates preferred parameter values.

Unfortunately, numerical data (in particular the most important transition temperatures and structural data) can not be searched and linked with the CAS Registry compound species at present. Nevertheless, the authors hope that the data presented will help scientists to cope with the abundance of information on superconductivity gathered within the last two decades.

**References and Information Sources**

Corresponding author: Dr. Werner Marx
Max Planck Institute for Solid State Research
Heisenbergstraße 1, D-70569 Stuttgart, Germany
E-mail: w.marx@fkf.mpg.de

E-mail Andreas Barth: andreas.barth@fiz-karlsruhe.de